\documentstyle[12pt] {article}
\newtheorem{theorem}{Theorem}[section]
\newtheorem{lemma}[theorem]{Lemma}
\newtheorem{definition}[theorem]{Definition}
\newtheorem{remark}[theorem]{Remark}

\newenvironment{proof}{\noindent{\bf Proof.\/}}{$~\Box$}

\newcommand{\beq}{\begin{equation}}
\newcommand{\eeq}{\end{equation}}
\newcommand{\beqa}{\begin{eqnarray}}
\newcommand{\eeqa}{\end{eqnarray}}
\newcommand{\ba}{\begin{array}}
\newcommand{\ea}{\end{array}}

\begin{document}

\begin{center}
\large
{\bf Non--Hyperbolic Dynamics: \\
a Family of Special Functions} 
\footnote{Presented to the 3rd International Summer School/Conference 
Let's Face Chaos Through Nonlinear Dynamics, 24 June -- 5 July, 1996, 
Maribor, Slovenia} \\
\vspace{0.25in}
\normalsize
{\bf Maria Morandi Cecchi}$^{(*)}$
\footnote{e--mail: mcecchi@math.unipd.it} 
and {\bf Luca Salasnich}$^{(*)(+)}$
\footnote{e--mail: salasnich@math.unipd.it} \\
\vspace{0.2in}
$^{(*)}$ Dipartimento di Matematica Pura ed Applicata \\
Universit\`a di Padova, Via Belzoni 7, I--35131 Padova, Italy \\
\vspace{0.2in}
$^{(+)}$ Istituto Nazionale di Fisica Nucleare \\
Sezione di Padova, Via Marzolo 8, I--35122 Padova, Italy \\
\end{center}

\vspace{0.3in}

{\bf Abstract.} We study the iterative dynamics of a family of special 
functions from ${\bf R}^2$ into ${\bf R}^2$ with a non--hyperbolic 
fixed point in the origin. The characterization by the eigenvalues is 
analyzed and discussed.


\vskip 0.5 truecm

\section{Introduction}
\par
The last twenty years have seen an explosion of interest in the study 
of nonlinear systems. Discrete dynamical systems, in particular 
one--dimensional, have been the subject of much active research, but 
only few results are known to be valid in any dimension. 
Moreover, very little is known so far on non--hyperbolic dynamics [1--4]. 
\par
In this paper we introduce some theorems for a class of non--hyperbolic fixed 
points on ${\bf R}^N$ and then analyze a family of functions $f_{\theta}$ 
on the plane which have a non--hyperbolic fixed point in the origin. 
The dynamical properties of the family near the fixed point, 
like the basin of attraction, are studied.  
Finally the limits of applicability of the characterization 
by the eigenvalues are discussed.  

\vskip 0.5 truecm

\section{Hyperbolic and non--hyperbolic points}
\par
The basic goal of the theory of dynamical systems is to understand the 
eventual or asymptotic behaviour of an iterative process. 
If the process is the iteration of a function $f$, then the theory hopes 
to understand the eventual behaviour of the points 
$x,f(x),f^2(x),...,f^n(x)$ as $n$ becomes large. 
Functions which determine dynamical systems are also called {\it maps}. 
This terminology connotes the geometric process of taking one point to 
another [1,2]. 
\par
We start with some elementary definitions for the discrete dynamics 
of maps on the plane. 

\begin{definition} 
Let $f:{\bf R}^N\to {\bf R}^N$. 
$f$ is a homeomorphism 
if is one--to--one, onto, and continuous, and $f^{-1}$ is also 
continuous. $f$ is a $C^r$--diffeomorphism if $f$ and $f^{-1}$ are 
homeomorphism of class $C^r$.
\end{definition}

\begin{definition} 
The forward orbit of 
$p\in {\bf R}^N$ for $f$ is the set of points $O^+(p)=\{p,f(p),f^2(p),...\}$. 
If $f$ is an homeomorphism, we define the full orbit of $p$ as 
the set of points $O(p)=\{f^n(p), \; \forall n\in Z\}$, 
and the backward orbit of $p$ 
as the set of points $O^-(p)=\{p,f^{-1}(p),f^{-2}(p),...\}$.
\end{definition}

\begin{definition} 
A point $p$ is a fixed point for $f$ 
if $f(p)=p$. We denote the set of fixed points by $Fix(f)$.
\end{definition}

\begin{definition} 
A point $p$ is a periodic point for $f$ 
with period $n$ if $f^n(p)=p$. The last positive $n$ for which 
$f^n(p)=p$ is called the prime period of $p$. 
We denote the set of periodic points of period $n$ by $Per_n(f)$.
\end{definition}

\begin{remark} 
A fixed point is a periodic point of prime period 1.
\end{remark}

\begin{definition} 
Let $p$ a periodic 
point of prime period $n$ for $f$, and $Df:{\bf R}^N\to {\bf R}^N\times 
{\bf R}^N$ the Jacobian of the first partial derivatives of $f$. 
The point $p$ is hyperbolic if $Df^n(p)$ does not have 
eigenvalues of unitary modulus. The point $p$ is non--hyperbolic if $Df^n(p)$ 
has at least one eigenvalue of unitary modulus.
\end{definition}

\begin{definition} 
Let $p$ a fixed point for $f$. It is called \\
i) attractor, if exists a neighbour $U$ of $p$ so that 
$$ 
\forall q\in U, \;\;\; \lim_{k\to \infty}f^k(q)=p ;
$$ 
\\
ii) source, if exists a neighbour $U$ of $p$ so that 
$$
\forall q\in U, \;\;\; \lim_{k\to \infty}f^{-k}(q)=p; 
$$
\\
iii) saddle point, in all the other cases.
\end{definition}

\begin{definition} 
Let $p$ an attractor for $f$. The basin of 
attraction of $p$ is the set of points
$$
w^{+}(p)=\{ q \in {\bf R}^N : \; \lim_{n\to \infty} f^n(q) = p \}.
$$
\end{definition}

\par
It is easy to characterize the properties of hyperbolic periodic points. 

\begin{theorem}
{\it Let $p$ a hyperbolic periodic point of prime period $n$ for $f$. 
Then the point $p$ is \\
(i) attractor, if all the eigenvalues of $Df^n(p)$ have modulus 
less than 1;\\
(ii) source, if all the eigenvalues of $Df^n(p)$ have modulus 
greater than 1;\\
(iii) saddle point, in all the other cases.}
\end{theorem}

\par
We do not proof this well--know theorem (see [1] or [4]) but we observe that 
they are valid also for periodic fixed point of prime 
period $n$, if one substitutes $f$ with $f^n$.
\par
A more difficult question is: How can we characterize the non--hyperbolic 
fixed point? {\it Id est}, 
what happen if some eigenvalues have unitary modulus?

\vskip 0.5 truecm

\section{Some theorems for non--hyperbolic fixed points}
\par
A partial answer to the question of the characterization of the 
non--hyperbolic fixed point is given by the following theorem. 
 
\begin{theorem}
{\it Let $f:{\bf R}^N\to {\bf R}^N$  
and $p\in {\bf R}^N$ a non--hyperbolic fixed point for $f$. 
Let $df(p):{\bf R}^N\to {\bf R}^N$ 
the differential of $f$ in $p$ so that 
$||df(p)||=1$ and $\forall q \in B_p/p$ $||df(q)||<1$, where $B_p$ is 
an open ball centered in $p$. Then $p$ is an attractor and 
$B_p \subset w^+(p)$, where $w^+(p)$ is the basin of attraction of $p$.} 
\end{theorem}

\begin{proof}
Let $q\in B_p$. From the definition of fixed point and the 
mean--value theorem we have
$$
||f(q)-p||=||f(q)-f(p)||\leq ||q - p|| sup\{ ||df(x)|| : \; x\in B_p^q \} ,
$$
where $B_p^q$ is the open ball centered in $p$ with radius $q$. 
We have that $sup\{ ||df(x)||: \; x\in B_p^q \} =||df(p)|| = 1$ and so
$$
\forall q\in B_p/p \;\;\; ||f(q)-p|| < ||q - p|| ,
$$
i.e the iterations of $q$ by $f$ remain in $B_p^q$ and their distance from 
$q$ decreases: $||f^{n}(q)-p||<||f^{n-1}(q)-p||$. As a consequence 
$\forall \epsilon >0$ exists a $m>0$ so that $||f^{m}(q)-p||<\epsilon$, 
thus the limit of $f^n(q)$ for $n\to \infty$ is $p$.  
\end{proof}

\vskip 0.4 truecm

\par
Now by using the theorem 3.1 we introduce another theorem 
that will be useful in the next section to study our family of special 
functions on the plane. First we proof the following lemma.

\begin{lemma} 
{\it Let $A$ a symmetric $2\times 2$ matrix. If 
its eigenvalues are less than one in modulus than $A$ 
has the property that
$$
\forall v\neq (0,0) \;\;\; || A v || < || v || , 
$$
where $||\;\;||$ is the Euclidean norm.} 
\end{lemma}

\begin{proof}
We know from the Cauchy-Schwarz inequality 
that for any compatible norm of the matrix $A$ 
$$
||A v||\leq ||A||\;||v||.
$$
Let $\lambda_1$ and $\lambda_2$ the eigenvalues of $A$ and 
$\rho (A)=max\{|\lambda_1|, \; |\lambda_2|\}$ the spectral radius of $A$. 
We have that $||A||=\sqrt{\rho (A^TA)}$ for the Euclidean norm by definition, 
but $A^TA=A^2$ because the matrix $A$ is symmetric. This means that 
$||A||=\sqrt{\rho(A^2)}=\rho (A)$, and because 
both the eigenvalues of $A$ are less than one in modulus 
$||A||=\rho (A) < 1$. In conclusion we obtain
$$
|| A v || \leq || A || \; || v || = \rho (A) \; ||v|| < || v || .
$$
This completes the proof.
\end{proof}

\begin{theorem} 
{\it Let $f:{\bf R}^2\to {\bf R}^2$  
and $(0,0)\in {\bf R}^2$ a non--hyperbolic fixed point for $f$, such as the 
eigenvalues of the Jacobian $Df(0,0)$ are $\lambda_1 ,\; 
\lambda_2 \in {\bf C}$; 
$|\lambda_1|=|\lambda_2|=1$. 
If exists an open ball $B_{(0,0)}$ centered in $(0,0)$ such as 
$\forall (x,y) \in B_{(0,0)}/(0,0)$ the eigenvalues of $Df(x,y)$ 
are less than 1 in modulus, then $(0,0)$ is an attractor and 
$B \subset w^+(0,0)$, where $w^+(0,0)$ is the basin of attraction of $(0,0)$.}
\end{theorem}

\begin{proof}
We want to show that $\forall q \in B_{(0,0)}$, 
$q\neq (0,0)$, $||f(q)||<||q||$. 
We consider the curve $\gamma (t): [0,1]\to B_{(0,0)}$, 
$\gamma (t) = t\cdot q$. 
The curve is monotonic, $f(\gamma (0))=0$, $f(\gamma (1)) =f(q)$, 
and $\gamma (t) \in B_{(0,0)}$ $\forall t\in [0,1]$. Then we have 
$$
||f(q)||=||\int_{[0,1]} f^{'}(\gamma (t)) dt || \leq 
\int_{]0,1[} ||Df(\gamma (t)) \gamma^{'}(t)|| dt 
$$
$$
< \int_{]0,1[} ||\gamma^{'}(t)|| dt = \int_{]0,1[} ||q|| dt = ||q||, 
$$
because the Jacobian $Df$ is a symmetric matrix. 
\end{proof}

\begin{remark} 
This theorem is true also if we eliminate 
a set of zero--measure from the domain of integration.
\end{remark}

\par
The theorem can be used to address the 
problem of the characterization of fixed points to the classical study 
of maxima and minima of the determinant of the Jacobian $Df(x,y)$ [5]. 

\begin{lemma} 
{\it Let $A$ a $2\times 2$ matrix. 
If the eigenvalues $\lambda_1, \lambda_2 \in {\bf C}/{\bf R}$ then 
$|\lambda_1|=|\lambda_2|=\sqrt{det(A)}$.}
\end{lemma}

\begin{proof} 
The determinant is given by  $det(A)= \lambda_1 \lambda_2 $. 
The non--real roots of any equation of degree 
2 are complex conjugate: ${\bar \lambda}_1=\lambda_2$. We have
$$
|\lambda_1 |^2 = \lambda_1 {\bar \lambda}_1 = 
\lambda_1 \lambda_2 = det(A). 
$$
In conclusion $|\lambda_1|=\sqrt{det(A)}$. The same proof can be applied 
to $\lambda_2$. 
\end{proof}

\begin{theorem} 
{\it Let $f:{\bf R}^2\to {\bf R}^2$  
and $(0,0)\in {\bf R}^2$ a non--hyperbolic fixed point for $f$ such as
the eigenvalues of the Jacobian $Df(0,0)$ are $\lambda_1 ,\; 
\lambda_2 \in {\bf C}/{\bf R} ,
\; |\lambda_1|=|\lambda_2|=1.$ 
If the function $det(Df(x,y))$ has a local maximum in $(0,0)$, 
then $(0,0)$ is an attractor.} 
\end{theorem}

\begin{proof} 
Because the eigenvalues of $Df(0,0)$ are complex 
but non--real, there is a neighbour of (0,0) in which such property 
is also true for continuity. If $det(Df(x,y))$ is a local maximum in 
(0,0) from the Lemma 3.5 we have that $det(Df(0,0))=1$ and that 
also the modulus of the eigenvalues is maximum in $(0,0)$. Now 
by using the theorem 3.3 we obtain the thesis. 
\end{proof}

\vskip 0.5 truecm

\section{A family of special functions: the $n$--polypous}
\par
Now we analyze a family of functions on the plane 
which has a non--hyperbolic fixed point in the origin. By using the 
theorems of the previous section we study analytically and numerically 
the dynamical properties of the family near the fixed point. 

\begin{figure}
\vskip 8. truecm
\caption{The dynamics of the 4--polypous (left) 
and the 12--polypous (right). 
Initial condition: $(1/2,1)$. Number of iterations: $5\times 10^3$.} 
\vskip 0.5 truecm
\end{figure}

\begin{definition} 
Let $f_{\theta}:{\bf R}^2\to {\bf R}^2$ such as 
$f_{\theta}(x,y)=(f^{(1)}_{\theta}(x,y),f^{(2)}_{\theta}(x,y))^T$, with 
$\theta \in [-2\pi ,2\pi ]$ and:
$$
f_{\theta}^{(1)}(x,y)=
x\cos{\theta} + y\sin{\theta} - {x^3\over 3} -{y^3\over 3}, 
$$
$$
f_{\theta}^{(2)}(x,y)= - x \sin{\theta} + y\cos{\theta} 
+ {x^3\over 3} - {y^3\over 3}.
$$
We call this family of functions $n$--polypous. 
\end{definition}

\par
To characterize the $n$--polypous we can calculate the Jacobian
\beq
Df_{\theta}(x,y)= 
\left( \begin{array}{cc} 
 \cos{\theta} - x^2 & \sin{\theta} - y^2 \\
-\sin{\theta} + x^2 & \cos{\theta} - y^2
\end{array} \right).
\eeq
The trace of $Df$ is given by 
\beq
Tr(Df_{\theta}(x,y))=2\cos{\theta} -(x^2+y^2) , 
\eeq
and the determinant 
\beq
det(Df_{\theta}(x,y))=2 x^2 y^2 -(\cos{\theta} + \sin{\theta})(x^2 +y^2) + 1. 
\eeq

\begin{figure}
\vskip 8. truecm
\caption{Basin of attraction $w^+(0,0)$. 
4--polypous (left) and 12--polypous (right). 
Lattice of $300\times 300$ initial conditions on the 
region $[-3,3]\times [-3,3]$. Number of iterations: $10^4$.} 
\vskip 0.5 truecm
\end{figure}

\begin{theorem} 
{\it The $n$--polypous has a non--hyperbolic 
fixed point in $(0,0)\in {\bf R}^2$. This fixed point is an attractor 
if $\theta \in \; ]-{\pi /4},{3\pi /4}[$, but $\theta \neq 0$.} 
\end{theorem}

\begin{proof} 
The point $(0,0) \in {\bf R}$ is a fixed point because 
$\forall \theta \in [-2\pi , 2\pi] \; f_{\theta}(0,0)=(0,0)$. 
From the equation (2) and (3) 
we obtain $\lambda_{1,2}(0,0)=(\cos{\theta}\pm i 
\sin{\theta})$. If $\theta \neq 0, \pi$, 
the eigenvalues are complex numbers but not real and $|\lambda_{1,2}|=1$. 
\par
We call $d(x,y)=det(Df_{\theta}(x,y))$ and obtain 
$$
\nabla d(x,y)= ( 4 x y^2 - 2 (\cos{\theta} + \sin{\theta})x , 
4 x^2 y - 2 (\cos{\theta} + \sin{\theta})y)^T ,
$$
and $\nabla d(0,0) = (0,0)$. Then 
$$
{\partial^2 d\over \partial x^2} = 4 y^2 - 2 (\cos{\theta} + \sin{\theta}), 
\;\;\;
{\partial^2 d\over \partial y^2} = 4 x^2 - 2 (\cos{\theta} + \sin{\theta}),
\;\;\;
{\partial^2 d\over \partial x \partial y} = 8 x y
$$
and the eigenvalues of the Hessian of $d$ in (0,0) are equal to 
$- 2 (\cos{\theta} + \sin{\theta})$. The Hessian Hd(0,0) is negative 
definite if $(\cos{\theta} + \sin{\theta}) > 0$, thus if 
$\theta \in ]-{\pi /4},{3\pi /4}[$. For these values of $\theta$ 
the origin is a local maximum and by using the theorem 3.6 
we have the thesis. 
\end{proof}

\begin{remark} 
Let $\theta = 2\pi /n$, $n\geq 3$. 
The dynamics of $f_{\theta}$ is attractive and, because $Df_{\theta}(0,0)$ is a 
rotation of $\theta$, it looks like a polypous with $n$ branches.
\end{remark}

\par
This remark justifies the name used for the family of these special functions. 
In Figure 1 we show the dynamics of $4$--polypous and the $12$--polypous 
for the same initial condition. 

\begin{figure}
\vskip 8. truecm
\caption{Region $S$ of the plane where the eigenvalues of 
the 12--polypous's Jacobian are less than one in modulus. 
Lattice of $300\times 300$ testing points on the region $[-3,3]\times [-3,3]$.} 
\vskip 0.5 truecm
\end{figure}

\par
The basin of attraction $w^+(0,0)$ of the origin for the 4--polypous and 
the 12--polypous is shown in Figure 2. This Figure is obtained by choosing 
a lattice of initial conditions on the plane and then by applying 
the definition 2.8. Our numerical calculations suggest that $w^+(0,0)$ 
is simply connected but it seems hard to find an analytical description for 
$w^+(0,0)$. 
\par
In Figure 3 we plot for the 12--polypous 
the region $S$ of the plane ${\bf R}^2$ 
where the eigenvalues of the Jacobian $Df_{\theta}(x,y)$ are in modulus 
less than one. This region is very different from the basin of attraction. 
In fact the theorems 3.1 and 3.3 are valid only for open balls $B_{(0,0)}$ 
centered in $(0,0)$; thus if $B_{(0,0)} \subset S$ than $B_{(0,0)} \subset 
w^+(0,0)$.  

\vskip 0.5 truecm
\section{Conclusions}
\par
We have studied some properties of a class of non--hyperbolic fixed 
points on ${\bf R}^N$ by introducing some theorems to characterize 
the attractors by the eigenvalues of the Jacobian matrix. 
We have analyzed the family of maps $f_{\theta}$, 
called $n$--polypous, showing that such maps have a non--hyperbolic 
fixed point in the origin. This fixed point is an attractor for many values 
of the parameter $\theta$ and numerical calculations 
suggest that its basin of attraction $w^+(0,0)$ is simply connected. 
The basin $w^+(0,0)$ changes by changing the values of the parameter 
$\theta$ and it is very different from the region $S$ of the plane 
where the eigenvalues of the Jacobian are in modulus less than one.  
\par
Our theorems, based on the study the eigenvalues of the Jacobian, 
are valid only for open balls centered in the origin of the fixed point. 
We thus conclude that the study of the eigenvalue of the Jacobian 
of the maps can be useful only for a local analysis of the fixed points.  

\vskip 0.5 truecm
\section*{References}

\parindent=0.pt

[1] J. Guckenheimer and P. Holmes, {\it Nonlinear Oscillations, 
Dynamical Systems, and Bifurcations of Vector Fields} (Springer, New York, 1983)

[2] R. Temam, {\it Infinite--Dimensional Dynamical Systems in Mechanics 
and Physics} (Springer, New York, 1988) 

[3] H. D. I. Abarbanel, M. I. Rabinovich and M. M. Sushchik, 
{\it Introduction to Nonlinear Dynamics for Physicists} (World Scientific, 
Singapore, 1993)

[4] A. H. Nayfeh and B. Balachandran, {\it Applied Nonlinear Dynamics} 
(Wiley, New York, 1995)

[5] P. Pellizari, {\it Dinamica non--iperbolica. Una famiglia di funzioni da 
${\bf R}^2$ a ${\bf R}^2$}, graduate thesis (advisor: Prof. 
M. Morandi Cecchi), Universit\`a di Padova (1991), 
unpublished. 

\end{document}